%% file: main.tex
\documentclass[sigconf,nonacm]{acmart}

\usepackage{url}            
\usepackage{booktabs}       
\usepackage{amsfonts}       
\usepackage{nicefrac}       
\usepackage{microtype}      
\usepackage{lipsum}		    
\usepackage{doi}
\usepackage{subcaption}
\usepackage{pifont}
\usepackage{colortbl}
\usepackage{float}
\usepackage{balance}
\usepackage{xspace}
\usepackage{acronym}
\usepackage{xcolor}
\usepackage{adjustbox}
\usepackage{siunitx}
\usepackage{tabularx}
\usepackage{enumitem}
\usepackage{multirow}
\usepackage{makecell}
\usepackage[subtle]{savetrees}

\usepackage{hyperref}   
\newcommand{\wcircle}[1]{\ding{\numexpr171 + #1}}
\newcommand{\bcircle}[1]{\ding{\numexpr181 + #1}}

\usepackage[most]{tcolorbox}

\usepackage[strict]{changepage} 
\usepackage{framed}            

\definecolor{promptbg}{HTML}{B7E0FF} 
\definecolor{promptborder}{HTML}{024CAA} 
\definecolor{answerbg}{rgb}{0.85,1,0.85} 
\definecolor{answerborder}{rgb}{0.0,0.6,0.30}

\newenvironment{answer}{
  
  \MakeFramed{\advance\hsize-\width\FrameRestore}
  \noindent\hspace{-4.55pt}
  \begin{adjustwidth}{}{7pt}
}{
  \end{adjustwidth}
  \endMakeFramed
}

\begin{document}

\title{Towards LLM-Enhanced Android Taint Analysis}

\author{Nicholas Miazzo}
\affiliation{
  \institution{University of Padova}
  \city{Padova}
  \country{Italy}
}
\email{nicholas.miazzo@math.unipd.it}

\author{Marco Alecci}
\affiliation{
  \institution{University of Luxembourg}
  \city{Luxembourg}
  \country{Luxembourg}
}
\email{marco.alecci@uni.lu}

\author{Jordan Samhi}
\affiliation{
  \institution{University of Luxembourg}
  \city{Luxembourg}
  \country{Luxembourg}
}
\email{jordan.samhi@uni.lu}

\author{Jacques Klein}
\affiliation{
  \institution{University of Luxembourg}
  \city{Luxembourg}
  \country{Luxembourg}
}
\email{jacques.klein@uni.lu}

\author{Eleonora Losiouk}
\affiliation{
  \institution{University of Padova}
  \city{Padova}
  \country{Italy}
}
\email{eleonora.losiouk@unipd.it}

\input{macro}

\pagestyle{plain}
\begin{abstract}
\input{00_Abstract.tex}

\end{abstract}

\maketitle

\input{01_Introduction.tex}

\input{02_Background}
\input{03_ExperimentalSetup}
\input{04_ExperimentalResults}
\input{05_Discussion}
\input{06_Limitations}
\input{07_RelatedWork}
\input{08_Conclusion}

\bibliographystyle{ACM-Reference-Format}
\bibliography{references}

\end{document}

%% file: macro.tex
\newcommand{\BestModel}{Gemini-3 Flash}
\newcommand{\BestModelx}{Gemini-3 Flash}
\newcommand{\BestFone}{0.96}
\newcommand{\BestFonex}{0.96}

\newcommand{\FDPrecision}{0.83}
\newcommand{\FDPrecisionx}{0.83}
\newcommand{\FDRecall}{0.42}
\newcommand{\FDRecallx}{0.42}
\newcommand{\FDFone}{0.55}
\newcommand{\FDFonex}{0.55}

\newcommand{\GeminiFlashPrecision}{\textbf{0.96}}
\newcommand{\GeminiFlashPrecisionx}{0.96}
\newcommand{\GeminiFlashRecall}{\textbf{0.95}}
\newcommand{\GeminiFlashRecallx}{0.95}
\newcommand{\GeminiFlashFone}{\textbf{0.96}}
\newcommand{\GeminiFlashFonex}{0.96}

\newcommand{\GeminiProPrecision}{XX}
\newcommand{\GeminiProPrecisionx}{XX}
\newcommand{\GeminiProRecall}{XX}
\newcommand{\GeminiProRecallx}{XX}
\newcommand{\GeminiProFone}{XX}
\newcommand{\GeminiProFonex}{XX}

\newcommand{\GLMPrecision}{0.15}
\newcommand{\GLMPrecisionx}{0.15}
\newcommand{\GLMRecall}{0.12}
\newcommand{\GLMRecallx}{0.12}
\newcommand{\GLMFone}{0.13}
\newcommand{\GLMFonex}{0.13}

\newcommand{\QwenPrecision}{0.87}
\newcommand{\QwenPrecisionx}{0.87}
\newcommand{\QwenRecall}{0.89}
\newcommand{\QwenRecallx}{0.89}
\newcommand{\QwenFone}{0.88}
\newcommand{\QwenFonex}{0.88}

\newcommand{\FDFoneAliasing}{0.50}
\newcommand{\FDFoneAliasingx}{0.50}
\newcommand{\FDFoneAndroidSpecific}{0.67}
\newcommand{\FDFoneAndroidSpecificx}{0.67}
\newcommand{\FDFoneArraysAndLists}{0.67}
\newcommand{\FDFoneArraysAndListsx}{0.67}
\newcommand{\FDFoneCallbacks}{0.50}
\newcommand{\FDFoneCallbacksx}{0.50}
\newcommand{\FDFoneDynamicLoading}{\textbf{0.50}}
\newcommand{\FDFoneDynamicLoadingx}{0.50}
\newcommand{\FDFoneEmulatorDetection}{0.94}
\newcommand{\FDFoneEmulatorDetectionx}{0.94}
\newcommand{\FDFoneFieldAndObject}{\textbf{1.00}}
\newcommand{\FDFoneFieldAndObjectx}{1.00}
\newcommand{\FDFoneGeneralJava}{0.47}
\newcommand{\FDFoneGeneralJavax}{0.47}
\newcommand{\FDFoneImplicitFlows}{0.00}
\newcommand{\FDFoneImplicitFlowsx}{0.00}
\newcommand{\FDFoneICC}{0.17}
\newcommand{\FDFoneICCx}{0.17}
\newcommand{\FDFoneLifecycle}{0.74}
\newcommand{\FDFoneLifecyclex}{0.74}
\newcommand{\FDFoneNative}{0.00}
\newcommand{\FDFoneNativex}{0.00}
\newcommand{\FDFoneReflection}{0.50}
\newcommand{\FDFoneReflectionx}{0.50}
\newcommand{\FDFoneReflectionCC}{0.00}
\newcommand{\FDFoneReflectionCCx}{0.00}
\newcommand{\FDFoneSelfModification}{0.00}
\newcommand{\FDFoneSelfModificationx}{0.00}
\newcommand{\FDFoneThreading}{\textbf{1.00}}
\newcommand{\FDFoneThreadingx}{1.00}
\newcommand{\FDFoneUnreachableCode}{\textbf{0.00}}
\newcommand{\FDFoneUnreachableCodex}{0.00}

\newcommand{\GeminiFlashFoneAliasing}{\textbf{0.67}}
\newcommand{\GeminiFlashFoneAliasingx}{0.67}
\newcommand{\GeminiFlashFoneAndroidSpecific}{\textbf{0.92}}
\newcommand{\GeminiFlashFoneAndroidSpecificx}{0.92}
\newcommand{\GeminiFlashFoneArraysAndLists}{\textbf{1.00}}
\newcommand{\GeminiFlashFoneArraysAndListsx}{1.00}
\newcommand{\GeminiFlashFoneCallbacks}{\textbf{0.98}}
\newcommand{\GeminiFlashFoneCallbacksx}{0.98}
\newcommand{\GeminiFlashFoneDynamicLoading}{0.00}
\newcommand{\GeminiFlashFoneDynamicLoadingx}{0.00}
\newcommand{\GeminiFlashFoneEmulatorDetection}{\textbf{1.00}}
\newcommand{\GeminiFlashFoneEmulatorDetectionx}{1.00}
\newcommand{\GeminiFlashFoneFieldAndObject}{\textbf{1.00}}
\newcommand{\GeminiFlashFoneFieldAndObjectx}{1.00}
\newcommand{\GeminiFlashFoneGeneralJava}{\textbf{0.98}}
\newcommand{\GeminiFlashFoneGeneralJavax}{0.98}
\newcommand{\GeminiFlashFoneImplicitFlows}{\textbf{0.94}}
\newcommand{\GeminiFlashFoneImplicitFlowsx}{0.94}
\newcommand{\GeminiFlashFoneICC}{0.95}
\newcommand{\GeminiFlashFoneICCx}{0.95}
\newcommand{\GeminiFlashFoneLifecycle}{\textbf{1.00}}
\newcommand{\GeminiFlashFoneLifecyclex}{1.00}
\newcommand{\GeminiFlashFoneNative}{\textbf{1.00}}
\newcommand{\GeminiFlashFoneNativex}{1.00}
\newcommand{\GeminiFlashFoneReflection}{\textbf{1.00}}
\newcommand{\GeminiFlashFoneReflectionx}{1.00}
\newcommand{\GeminiFlashFoneReflectionCC}{\textbf{1.00}}
\newcommand{\GeminiFlashFoneReflectionCCx}{1.00}
\newcommand{\GeminiFlashFoneSelfModification}{\textbf{0.86}}
\newcommand{\GeminiFlashFoneSelfModificationx}{0.86}
\newcommand{\GeminiFlashFoneThreading}{\textbf{1.00}}
\newcommand{\GeminiFlashFoneThreadingx}{1.00}
\newcommand{\GeminiFlashFoneUnreachableCode}{\textbf{0.00}}
\newcommand{\GeminiFlashFoneUnreachableCodex}{0.00}

\newcommand{\GeminiProFoneAliasing}{XX}
\newcommand{\GeminiProFoneAliasingx}{XX}
\newcommand{\GeminiProFoneAndroidSpecific}{XX}
\newcommand{\GeminiProFoneAndroidSpecificx}{XX}
\newcommand{\GeminiProFoneArraysAndLists}{XX}
\newcommand{\GeminiProFoneArraysAndListsx}{XX}
\newcommand{\GeminiProFoneCallbacks}{XX}
\newcommand{\GeminiProFoneCallbacksx}{XX}
\newcommand{\GeminiProFoneDynamicLoading}{XX}
\newcommand{\GeminiProFoneDynamicLoadingx}{XX}
\newcommand{\GeminiProFoneEmulatorDetection}{XX}
\newcommand{\GeminiProFoneEmulatorDetectionx}{XX}
\newcommand{\GeminiProFoneFieldAndObject}{XX}
\newcommand{\GeminiProFoneFieldAndObjectx}{XX}
\newcommand{\GeminiProFoneGeneralJava}{XX}
\newcommand{\GeminiProFoneGeneralJavax}{XX}
\newcommand{\GeminiProFoneImplicitFlows}{XX}
\newcommand{\GeminiProFoneImplicitFlowsx}{XX}
\newcommand{\GeminiProFoneICC}{XX}
\newcommand{\GeminiProFoneICCx}{XX}
\newcommand{\GeminiProFoneLifecycle}{XX}
\newcommand{\GeminiProFoneLifecyclex}{XX}
\newcommand{\GeminiProFoneNative}{XX}
\newcommand{\GeminiProFoneNativex}{XX}
\newcommand{\GeminiProFoneReflection}{XX}
\newcommand{\GeminiProFoneReflectionx}{XX}
\newcommand{\GeminiProFoneReflectionCC}{XX}
\newcommand{\GeminiProFoneReflectionCCx}{XX}
\newcommand{\GeminiProFoneSelfModification}{XX}
\newcommand{\GeminiProFoneSelfModificationx}{XX}
\newcommand{\GeminiProFoneThreading}{XX}
\newcommand{\GeminiProFoneThreadingx}{XX}
\newcommand{\GeminiProFoneUnreachableCode}{XX}
\newcommand{\GeminiProFoneUnreachableCodex}{XX}

\newcommand{\GLMFoneAliasing}{0.00}
\newcommand{\GLMFoneAliasingx}{0.00}
\newcommand{\GLMFoneAndroidSpecific}{0.29}
\newcommand{\GLMFoneAndroidSpecificx}{0.29}
\newcommand{\GLMFoneArraysAndLists}{0.57}
\newcommand{\GLMFoneArraysAndListsx}{0.57}
\newcommand{\GLMFoneCallbacks}{0.00}
\newcommand{\GLMFoneCallbacksx}{0.00}
\newcommand{\GLMFoneDynamicLoading}{0.00}
\newcommand{\GLMFoneDynamicLoadingx}{0.00}
\newcommand{\GLMFoneEmulatorDetection}{0.34}
\newcommand{\GLMFoneEmulatorDetectionx}{0.34}
\newcommand{\GLMFoneFieldAndObject}{0.00}
\newcommand{\GLMFoneFieldAndObjectx}{0.00}
\newcommand{\GLMFoneGeneralJava}{0.27}
\newcommand{\GLMFoneGeneralJavax}{0.27}
\newcommand{\GLMFoneImplicitFlows}{0.00}
\newcommand{\GLMFoneImplicitFlowsx}{0.00}
\newcommand{\GLMFoneICC}{0.05}
\newcommand{\GLMFoneICCx}{0.05}
\newcommand{\GLMFoneLifecycle}{0.05}
\newcommand{\GLMFoneLifecyclex}{0.05}
\newcommand{\GLMFoneNative}{0.67}
\newcommand{\GLMFoneNativex}{0.67}
\newcommand{\GLMFoneReflection}{0.00}
\newcommand{\GLMFoneReflectionx}{0.00}
\newcommand{\GLMFoneReflectionCC}{0.00}
\newcommand{\GLMFoneReflectionCCx}{0.00}
\newcommand{\GLMFoneSelfModification}{0.00}
\newcommand{\GLMFoneSelfModificationx}{0.00}
\newcommand{\GLMFoneThreading}{0.18}
\newcommand{\GLMFoneThreadingx}{0.18}
\newcommand{\GLMFoneUnreachableCode}{\textbf{0.00}}
\newcommand{\GLMFoneUnreachableCodex}{0.00}

\newcommand{\QwenFoneAliasing}{0.50}
\newcommand{\QwenFoneAliasingx}{0.50}
\newcommand{\QwenFoneAndroidSpecific}{0.80}
\newcommand{\QwenFoneAndroidSpecificx}{0.80}
\newcommand{\QwenFoneArraysAndLists}{\textbf{1.00}}
\newcommand{\QwenFoneArraysAndListsx}{1.00}
\newcommand{\QwenFoneCallbacks}{0.83}
\newcommand{\QwenFoneCallbacksx}{0.83}
\newcommand{\QwenFoneDynamicLoading}{0.00}
\newcommand{\QwenFoneDynamicLoadingx}{0.00}
\newcommand{\QwenFoneEmulatorDetection}{\textbf{1.00}}
\newcommand{\QwenFoneEmulatorDetectionx}{1.00}
\newcommand{\QwenFoneFieldAndObject}{\textbf{1.00}}
\newcommand{\QwenFoneFieldAndObjectx}{1.00}
\newcommand{\QwenFoneGeneralJava}{0.90}
\newcommand{\QwenFoneGeneralJavax}{0.90}
\newcommand{\QwenFoneImplicitFlows}{0.22}
\newcommand{\QwenFoneImplicitFlowsx}{0.22}
\newcommand{\QwenFoneICC}{\textbf{0.97}}
\newcommand{\QwenFoneICCx}{0.97}
\newcommand{\QwenFoneLifecycle}{\textbf{1.00}}
\newcommand{\QwenFoneLifecyclex}{1.00}
\newcommand{\QwenFoneNative}{\textbf{1.00}}
\newcommand{\QwenFoneNativex}{1.00}
\newcommand{\QwenFoneReflection}{\textbf{1.00}}
\newcommand{\QwenFoneReflectionx}{1.00}
\newcommand{\QwenFoneReflectionCC}{0.96}
\newcommand{\QwenFoneReflectionCCx}{0.96}
\newcommand{\QwenFoneSelfModification}{0.00}
\newcommand{\QwenFoneSelfModificationx}{0.00}
\newcommand{\QwenFoneThreading}{\textbf{1.00}}
\newcommand{\QwenFoneThreadingx}{1.00}
\newcommand{\QwenFoneUnreachableCode}{\textbf{0.00}}
\newcommand{\QwenFoneUnreachableCodex}{0.00}

%% file: 00_Abstract.tex
Taint analysis is a fundamental technique for detecting sensitive data leaks in Android apps. However, traditional static tools, such as FlowDroid, still face well-known challenges due to the complexity of accurately modeling the Android framework. In this paper, we investigate whether off-the-shelf Large Language Models (LLMs) can effectively reason about taint flows in Android apps. Our preliminary approach relies on an agentic interaction strategy, enabling the LLM to iteratively explore code and reason about data flows.

We conduct an initial evaluation on the \textsc{DroidBench} benchmark against FlowDroid, where our approach outperforms the baseline: \BestModelx{} achieves an F1-score of \BestFonex{}, compared to \FDFonex{} for FlowDroid. In particular, we observe improvements in challenging categories such as inter-component communication (\GeminiFlashFoneICCx{} vs. \FDFoneICCx{}), implicit flows (\GeminiFlashFoneImplicitFlowsx{} vs. \FDFoneImplicitFlowsx{}), and reflection (\GeminiFlashFoneReflectionx{} vs. \FDFoneReflectionx{}), where FlowDroid typically struggles. On a small set of real-world apps, the LLM-based approach also identifies additional potential data leaks not reported by FlowDroid. These preliminary findings suggest that LLM reasoning may effectively complement traditional static taint analysis, motivating future research on hybrid LLM-enhanced taint analysis pipelines.

%% file: 01_Introduction.tex
\section{Introduction}
\label{sec:introduction}

Taint analysis is a fundamental technique for detecting sensitive data leaks in Android applications (apps), enabling the identification of flows from privacy-critical sources (e.g., device identifiers, location data, credentials) to potentially unsafe sinks such as network interfaces or logs. Over the past decade, static analysis tools such as FlowDroid~\cite{arzt2014flowdroid} have become widely adopted due to their ability to perform precise, context-, flow-, and lifecycle-aware analysis of Android apps. However, these approaches rely on carefully engineered models of the Android framework and struggle with well-known challenges, including reflection, dynamic code loading, inter-component communication, and incomplete code~\cite{luo2022taintbench,samhi2024call,zhang2021analyzing, zhang2021condysta, samhi2025you}.
In parallel, Large Language Models (LLMs) are increasingly being adopted across a wide range of software engineering tasks, including bug detection and fixing~\cite{feng2024prompting, li2024enhancing, bouzenia2024repairagent, kang2023large}, testing~\cite{liu2024make, huang2024crashtranslator, chen2024chatunitest, schafer2023empirical}, vulnerability detection~\cite{zhao2025apppoet, qian2025lamd, sun2024gptscan, guo2024outside, sun2025raml}, and other applications~\cite{khare2023understanding, pei2023can, ma2023lms, sun2023automatic, zhong2024can, alecci2025toward}. Their ability to understand and reason over both natural and programming languages enables them to bridge gaps traditionally filled by domain-specific models and heuristics.

As LLM capabilities continue to advance, a key question arises:
\emph{Can an off-the-shelf LLM, without task-specific training or handcrafted framework models, effectively reason about taint flows in Android apps?} In this paper, we investigate this question through a preliminary study on the use of LLM-driven reasoning for Android taint analysis. Instead of constructing explicit models of the Android framework, we enable an LLM to iteratively explore decompiled apps and reason about potential data flows via a Model Context Protocol (MCP)-based toolchain. In this setting, the LLM acts as an analysis agent, dynamically navigating code and forming hypotheses about source-to-sink relationships. Our goal is not to propose a production-ready taint analyzer, but to assess whether agentic LLM reasoning can potentially complement traditional static taint analysis approaches. To this end, we conduct an initial evaluation on the \textsc{DroidBench} benchmark and a small set of real-world apps, comparing against the state-of-the-art FlowDroid.

Our findings provide early evidence that LLM-based reasoning can identify taint flows, often achieving higher recall in scenarios involving complex behaviors (e.g., ICC, reflection, and implicit flows), while showing comparable performance in more standard cases. This indicates that LLMs are particularly useful when traditional modeling becomes difficult, but are not always necessary. Rather than applying LLMs indiscriminately, these findings point to a hybrid approach: traditional analyzers handle common cases efficiently, while LLMs could be selectively applied to harder scenarios. Such a design could improve overall effectiveness while also helping balance computational cost and scalability. As future work, we plan to explore this direction by designing more advanced hybrid analysis pipelines that integrate static analysis and LLM-based reasoning, as well as extending our evaluation to a larger set of real-world apps.

\noindent
\textbf{Contributions.} This paper makes the following contributions:
\begin{itemize}[leftmargin=*]
    \item We explore a novel perspective on Android taint analysis, investigating whether LLM can effectively reason about taint flows in Android apps.
    \item We design an LLM-driven taint analysis approach that combines decompiled code with MCP-enabled tool interaction.
    \item We provide a preliminary empirical study showing that LLM-based reasoning can complement traditional analyzers such as FlowDroid, particularly in challenging scenarios, motivating further studies on hybrid taint analysis approaches combining static analysis and agentic LLM reasoning.
\end{itemize}

%% file: 02_Background.tex
\section{Background}
\label{sec:background}
In this section, we briefly describe two concepts necessary to understand the remainder of the paper.

\noindent
\textbf{Taint Analysis for Android}.
Taint analysis tracks the flow of data from predefined \emph{sources} to potentially dangerous \emph{sinks}. In the Android ecosystem, this task is complicated by the event-driven execution model, component lifecycles, inter-component communication (ICC), and extensive framework APIs. FlowDroid~\cite{arzt2014flowdroid} is a widely used static taint analysis tool for Android that achieves high precision by modeling Android lifecycles and callbacks. While effective on benchmarks, FlowDroid depends on manually maintained models and may miss flows involving reflection, dynamically registered callbacks, or incomplete code. In this paper, we explore whether LLMs can achieve comparable results without relying on explicit framework modeling or task-specific training.

\noindent
\textbf{Model Context Protocol (MCP).}
LLMs usually operate in a closed setting and cannot directly interact with external tools or program representations. MCP (Model Context Protocol)~\cite{mcp} is an open standard that enables LLMs to interact with external tools and data sources through a unified interface, allowing models to query, navigate, and reason over structured information. In practice, MCP exposes external capabilities as callable tools that the LLM can invoke during its reasoning process, enabling iterative exploration of complex artifacts such as full Android apps.

%% file: 03_ExperimentalSetup.tex
\section{Experimental Setup}
\label{sec:experimentalsetup}

In this section, we present our approach and implementation details.

\begin{figure}[ht]
    \centering
    \begin{adjustbox}{width=\linewidth}
   \includegraphics{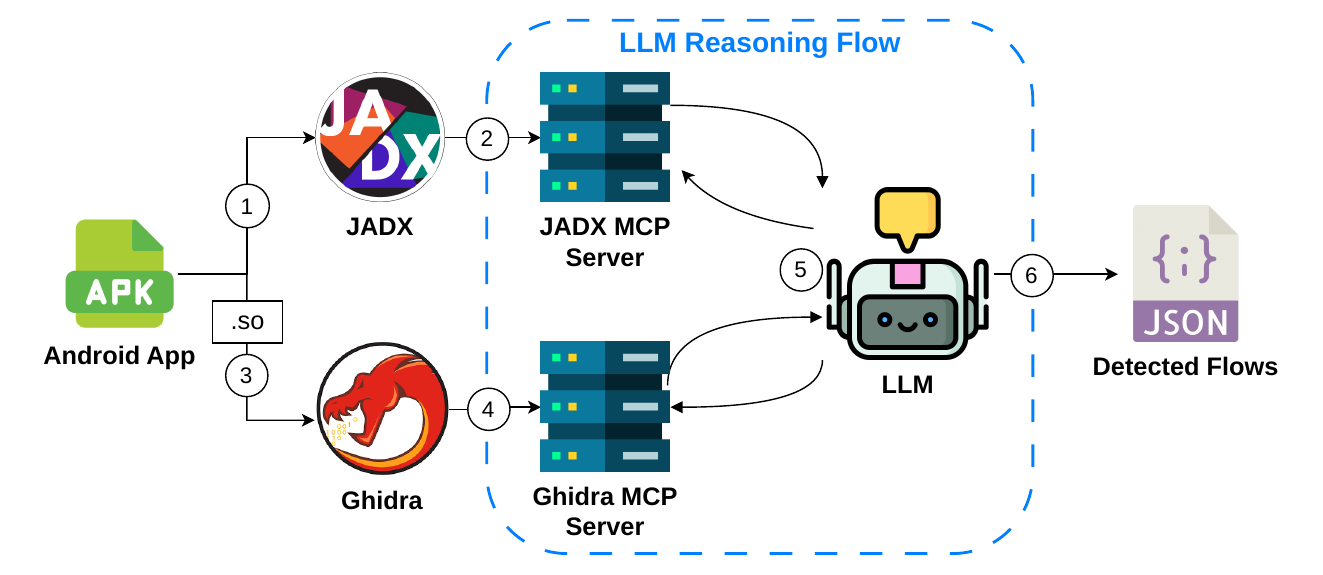}
    \end{adjustbox}
    \caption{Approach Overview}
    \label{fig:approachOVerview}
\end{figure}

\noindent
\textbf{Approach Overview.}
Figure~\ref{fig:approachOVerview} presents an overview of our approach. The core idea is to use an LLM as an analysis agent that interacts with MCP-enabled reverse-engineering tools to explore Android apps and reason about potential taint flows. Rather than relying on manually engineered Android framework models, the LLM dynamically navigates the codebase to infer source-to-sink relationships. To support this process, we rely on \texttt{JADX-MCP}~\cite{jadx-mcp} and \texttt{Ghidra-MCP}~\cite{ghidra-mcp}, which expose decompilation and program-analysis capabilities (e.g., method retrieval, cross-references, and class hierarchies) as MCP tools.

The analysis workflow proceeds as follows. First, the Android APK is loaded into JADX for decompilation \wcircle{1}. The associated JADX MCP server \wcircle{2} exposes APIs that allow the LLM to query and navigate the recovered application structure (e.g., \texttt{get\_\-method\_\-by\_name()}). If the app contains native libraries, the extracted \texttt{.so} binary is loaded into Ghidra for analysis \wcircle{3}. Through the Ghidra MCP server \wcircle{4}, the LLM can similarly inspect native functions, references, and low-level program structures. Using these MCP-enabled tools, the LLM iteratively explores the application codebase \wcircle{5}, traces the propagation of sensitive information, and reasons about potential taint flows across components and languages. Finally, all identified flows are collected and serialized into a structured JSON report \wcircle{6}. The full prompts used in our experiments are available in our repository.

To counter the non-determinism of LLMs, which can produce hallucinations and spurious outputs, each app is analyzed $N$ times using the same prompt, and a detected flow is considered valid only if it appears in at least $\lfloor N/2 \rfloor + 1$ analyses. The value of $N$ can be customized depending on the use case and available resources.

\noindent
\textbf{Implementation Details.}
For this project, we combine proprietary models with locally hosted open-source models, using agentic CLIs in both cases. This choice addresses the two main deployment scenarios, covering both convenience-oriented closed-source services and self-hosted solutions for cost containment and confidentiality. For proprietary models, we use \textbf{Gemini 3 Flash}~\cite{gemini3}, while for the open-source model we select \textbf{Qwen3.5-27B}~\cite{qwen35}. We orchestrate these models through \textbf{Claude Code}~\cite{claudecodelocal} and \textbf{Gemini CLI}~\cite{geminicli}. Additional implementation details and configurations are available in the accompanying repository.

Regarding the number of analyses (the $N$ parameter), we performed 5 runs for open-source models and 3 for closed-source models (due to API cost constraints) in the baseline evaluation (see RQ1 and RQ2), and 5 for real-world apps assessment (see RQ3).

\noindent
\textbf{Comparison with FlowDroid.}
The LLM-based approach and FlowDroid are executed independently. During the analysis process, the LLM does not receive any information derived from FlowDroid outputs, detected flows, or taint specifications. For the comparison, we run FlowDroid using its default configuration, including its default source/sink definitions, which are available on their repo~\cite{FlowDroidToolRepo}. We intentionally adopt this default setup to reflect a common usage scenario in which FlowDroid is employed as a black-box taint analysis tool. We further discuss this choice and its implications in Section~\ref{sec:limitations}.

%% file: 04_ExperimentalResults.tex
\section{Experimental Results}
\label{sec:results}

In this section, we present the results of our preliminary evaluation, structured around the following research questions (RQs):
\begin{itemize}[leftmargin=*]
\item \textbf{RQ1:} How does the LLM-based taint analysis approach compare to FlowDroid on \textsc{DroidBench}?
\item \textbf{RQ2:} How does the LLM-based approach perform across the different categories of \textsc{DroidBench} compared to FlowDroid?
\item \textbf{RQ3:} How does the LLM-based approach perform when applied to real-world Android applications?
\end{itemize}

\subsection{RQ1: Overall Performance on \textsc{DroidBench}}
\label{sec:rq1}

To answer RQ1, we compare our LLM-based approach against FlowDroid on \textsc{DroidBench}~\cite{arzt2014flowdroid}, which is a widely used benchmark for Android taint analysis, consisting (in version 3.0) of 190 test cases across 19 categories, covering challenges such as lifecycle modeling, asynchronous callbacks, and UI interactions \footnote{Inter-App Communication excluded since neither approach supports it.}. We evaluate precision, recall, and F1-score against the ground truth. Table~\ref{tab:rq1-overall} reports the aggregated results.

\input{table1}

FlowDroid achieves an F1-score of \FDFonex{}, reflecting its conservative analysis strategy and relatively low recall (\FDRecallx{}). While its precision remains high (\FDPrecisionx{}), it fails to detect a substantial portion of known taint flows, leading to many false negatives.
In contrast, our LLM-based approach outperforms FlowDroid in terms of F1-score. In particular, \BestModelx{} achieves the best overall performance with an F1-score of \BestFone{}, substantially improving over the FlowDroid baseline. This gain is primarily driven by a large increase in recall (up to \GeminiFlashRecallx{}), while maintaining competitive precision. Qwen3.5-27B also improves over FlowDroid, reaching an F1-score of \QwenFonex{}. 
Overall, these results provide preliminary evidence that LLMs can improve taint-flow detection on \textsc{DroidBench}, especially by reducing false negatives, which we further analyze in RQ2 (Section~\ref{sec:rq2}).

\begin{answer}
\textbf{Answer to RQ1:} LLM-based taint analysis can outperform FlowDroid on \textsc{DroidBench}, primarily due to substantially higher recall.
\end{answer}

\subsection{RQ2: \textsc{DroidBench} Categories}
\label{sec:rq2}

To answer RQ2, we further analyze the evaluation results from RQ1 by disaggregating them across the benchmark categories defined in \textsc{DroidBench}. The detailed definitions of these categories are available on the official \textsc{DroidBench} GitHub page~\cite{DroidBench}. Table~\ref{tab:rq2categories} reports the corresponding F1-scores for each category.

\input{table2.tex}

As already observed in RQ1, the effectiveness of LLM-based taint analysis varies significantly across models, with some models outperforming FlowDroid across several \textsc{DroidBench} categories, while others remain less effective.
In particular, substantial improvements are observed in categories such as \emph{InterComponentCommunication} (ICC), \emph{ImplicitFlows}, \emph{Reflection}, and \emph{Native}, where FlowDroid exhibits limited or no recall. These categories are known to be particularly challenging for traditional static analysis due to complex control flow, dynamic dispatch, incomplete framework modeling, and native code boundaries. The results suggest that LLM-based approaches can better reason about these complex behaviors by leveraging semantic understanding of decompiled code, enabling the identification of taint propagation patterns that are missed by the baseline. At the same time, categories such as \emph{DynamicLoading} remain challenging for both traditional and LLM-based approaches, while others exhibit similar performance. Overall, the category-level analysis suggests that LLMs are particularly promising as a complementary technique for challenging analysis scenarios, as further discussed in Section~\ref{sec:discussion}.

\begin{answer}
\textbf{Answer to RQ2:} LLM-based approaches outperform FlowDroid in most \textsc{DroidBench} categories, particularly those involving ICC, implicit flows, reflection, and native code.
\end{answer}

\subsection{RQ3: Performance on Real-world Apps}
\label{sec:rq3}
To answer RQ3, we evaluate both FlowDroid and our LLM-based approach on a small set of real-world Android apps. Given the exploratory nature of this study and the absence of ground truth for real-world taint flows, we analyzed 5 apps randomly selected from AndroZoo~\cite{androzoo2016}, drawn from Google Play within the last 5 years. We deliberately limit the sample size to enable thorough manual inspection of each reported flow, which is necessary to assess its validity. Accordingly, our goal is not statistical generalization, but a qualitative, first assessment of whether LLM-based reasoning can uncover additional real data leaks beyond those detected by FlowDroid. 

\begin{figure}[H]
    \centering
    \begin{adjustbox}{width=\linewidth}
   \includegraphics{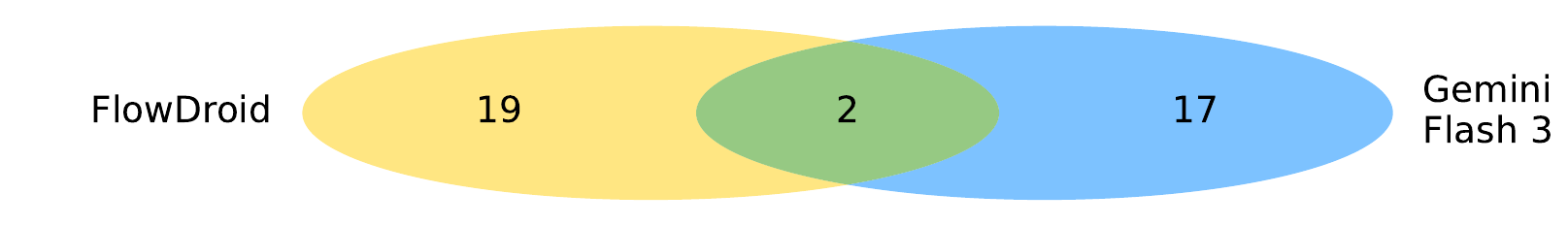}
    \end{adjustbox}
    \caption{Overlap between FlowDroid and Gemini 3 Flash.}
    \label{fig:rq3_venn}
\end{figure}

As discussed in Section~\ref{sec:experimentalsetup}, our baseline comparison uses the default FlowDroid configuration, whose default source/sink definitions are sufficient for \textsc{DroidBench}. However, for the real-world app evaluation, we also extended FlowDroid’s source/sink definitions to include all sources and sinks identified by the LLM-based approach. This was done to ensure a fair comparison, preventing FlowDroid from missing flows solely because the corresponding sources or sinks were absent from its default configuration. Figure~\ref{fig:rq3_venn} presents a Venn diagram illustrating the overlap between the taint flows detected by FlowDroid and those identified by our approach. Due to the cost and computational overhead of manually validating flows across multiple runs, we conduct the real-world evaluation only with the best-performing LLM configuration from RQ1 and RQ2, i.e., Gemini 3 Flash.

Unlike DroidBench, real-world apps do not provide ground truth for all taint flows, making recall extremely challenging to compute without exhaustive manual reverse engineering of the entire apps. Therefore, we focus on the 17 additional flows reported by the LLM-based approach (i.e., the light blue region in Figure~\ref{fig:rq3_venn}).  We do not assume that all flows reported by FlowDroid are true positives; however, our objective is to evaluate the ability of the LLM-based approach to identify previously undetected flows beyond those already reported by the baseline analyzer.
Two annotators with expertise in Android taint analysis independently inspected these flows by manually reverse-engineering the apps to determine whether each corresponded to a true or false positive. In cases of disagreement, the annotators discussed their findings until reaching a consensus. It is important to note that our goal is not to assess whether the flows are malicious (which is beyond the scope of this paper), but rather to determine whether the reported source-to-sink flows actually exist in the analyzed apps. Our analysis reveals that 16 out of the 17 additional flows correspond to true positives, while one was revealed to be a hallucination from the LLM. These findings indicate that LLM-based approaches can uncover previously undetected data leaks in real-world apps, thereby complementing traditional static analysis tools. For example, in one analyzed app, the source method itself is retrieved via reflection. In this scenario, FlowDroid cannot even start the corresponding taint analysis, since the reflective invocation prevents the source method from being resolved during the source lookup phase. In contrast, the LLM-based approach can still reason about the flow through iterative code exploration. At the same time, the LLM-based approach also misses several flows detected by FlowDroid, highlighting that the two approaches capture partially different classes of taint flows. This observation motivates further investigation with a specific focus on hybrid approaches that combine traditional static analysis with LLM-based reasoning, as discussed in Section~\ref{sec:discussion}.

\noindent
\textbf{Considerations on Non-determinism.}
To mitigate non- determinism, flows are retained only if reported in the majority of runs (see Section~\ref{sec:experimentalsetup}). Manual inspection nevertheless revealed additional true positives below this threshold, suggesting that increasing the number of runs may further improve coverage and uncover additional valid taint flows.

\begin{answer}
\textbf{Answer to RQ3:} The LLM-based approach is able to identify additional true data leaks in real-world apps beyond those detected by FlowDroid, although challenges remain in terms of reliability, evaluation, and completeness.
\end{answer}

%% file: table1.tex
\begin{table}[ht]
\centering
\caption{Overall performance on \textsc{DroidBench}.}
\label{tab:rq1-overall}
\small
\begin{adjustbox}{width=0.7\columnwidth}
\begin{tabular}{lccc}
\toprule
\textbf{Metric} & \textbf{FlowDroid} & \textbf{\makecell{Gemini-3 \\ Flash}} & \textbf{\makecell{Qwen3.5 \\ 27B}} \\
\midrule
Precision & \FDPrecision & \GeminiFlashPrecision & \QwenPrecision \\
Recall & \FDRecall & \GeminiFlashRecall & \QwenRecall \\
F1-score & \FDFone & \GeminiFlashFone & \QwenFone \\
\bottomrule
\end{tabular}
\end{adjustbox}
\end{table}

%% file: table2.tex
\begin{table}[ht]
\centering
\caption{Category-wise performance (F1) on \textsc{DroidBench}.}
\label{tab:rq2categories}
\begin{adjustbox}{width=0.9\columnwidth}
\small
\begin{tabular}{lccc}
\toprule
\textbf{Category} & \textbf{FlowDroid} & \textbf{\makecell{Gemini-3 \\ Flash}} & \textbf{\makecell{Qwen3.5 \\ 27B}} \\
\midrule
Aliasing & \FDFoneAliasing & \GeminiFlashFoneAliasing & \QwenFoneAliasing \\
AndroidSpecific & \FDFoneAndroidSpecific & \GeminiFlashFoneAndroidSpecific & \QwenFoneAndroidSpecific \\
ArraysAndLists & \FDFoneArraysAndLists & \GeminiFlashFoneArraysAndLists & \QwenFoneArraysAndLists \\
Callbacks & \FDFoneCallbacks & \GeminiFlashFoneCallbacks & \QwenFoneCallbacks \\
DynamicLoading & \FDFoneDynamicLoading & \GeminiFlashFoneDynamicLoading & \QwenFoneDynamicLoading \\
EmulatorDetection & \FDFoneEmulatorDetection & \GeminiFlashFoneEmulatorDetection & \QwenFoneEmulatorDetection \\
FieldAndObject Sensitivity & \FDFoneFieldAndObject & \GeminiFlashFoneFieldAndObject & \QwenFoneFieldAndObject \\
GeneralJava & \FDFoneGeneralJava & \GeminiFlashFoneGeneralJava & \QwenFoneGeneralJava \\
ImplicitFlows & \FDFoneImplicitFlows & \GeminiFlashFoneImplicitFlows & \QwenFoneImplicitFlows \\
ICC & \FDFoneICC & \GeminiFlashFoneICC & \QwenFoneICC \\
Lifecycle & \FDFoneLifecycle & \GeminiFlashFoneLifecycle & \QwenFoneLifecycle \\
Native & \FDFoneNative & \GeminiFlashFoneNative & \QwenFoneNative \\
Reflection & \FDFoneReflection & \GeminiFlashFoneReflection & \QwenFoneReflection \\
Reflection\_CC & \FDFoneReflectionCC & \GeminiFlashFoneReflectionCC & \QwenFoneReflectionCC \\
SelfModification & \FDFoneSelfModification & \GeminiFlashFoneSelfModification & \QwenFoneSelfModification \\
Threading & \FDFoneThreading & \GeminiFlashFoneThreading & \QwenFoneThreading \\
UnreachableCode & \FDFoneUnreachableCode & \GeminiFlashFoneUnreachableCode & \QwenFoneUnreachableCode \\

\bottomrule
\end{tabular}
\end{adjustbox}
\end{table}

%% file: 05_Discussion.tex
\section{Discussion and Research Agenda}
\label{sec:discussion}

Several important considerations arise from our results:

\noindent
\bcircle{1} \textbf{LLMs without framework modeling.}
Our results suggest that LLMs can reason about Android taint flows without any explicit training or handcrafted knowledge about the Android framework. Despite this, it is still able to identify non-trivial taint flows. This suggests that LLMs can potentially lower the engineering effort typically required by traditional static analyzers. At the same time, the effectiveness of the approach strongly depends on the underlying LLM, potentially introducing trade-offs between analysis quality, scalability, and computational cost.

\noindent
\bcircle{2} \textbf{Complementarity with traditional analysis.}
LLMs appear particularly effective in challenging scenarios such as ICC, implicit flows, reflection, and native code, where traditional analyzers often struggle. However, they may be unnecessary or overly expensive for simpler taint propagation cases. Therefore, one potential direction could be hybrid approaches in which traditional analyzers handle common cases, while LLMs are selectively applied only to more challenging analysis scenarios.

\noindent
\bcircle{3} \textbf{Challenges in evaluation.}
While benchmarks such as DroidBench provide ground truth, evaluating taint analysis on real-world apps remains challenging due to the lack of labeled datasets and the need for expensive manual validation.

\noindent\textbf{Research Agenda.}
This work represents a preliminary step towards understanding the role of LLMs in taint analysis. We outline the following directions for future work:
\begin{itemize}[leftmargin=*]
\item \textbf{Hybrid and explainable analysis.}
We plan to investigate hybrid pipelines combining static analysis and LLM-based reasoning, where LLMs are selectively triggered only for challenging scenarios. In addition, we aim to improve the explainability of detected taint flows for human analysts and developers.
\item \textbf{Larger-scale evaluation.}
We aim to extend the evaluation to a broader set of real-world apps, additional LLMs, and more advanced prompting strategies to better assess robustness and generalizability.
\item \textbf{Cost-aware analysis.}
We plan to systematically study the trade-offs between effectiveness, scalability, and computational cost, including strategies for selectively invoking stronger LLMs only for challenging analysis scenarios.
\end{itemize}

%% file: 06_Limitations.tex
\section{Limitations}
\label{sec:limitations}
In this section, we discuss the main limitations of our study.

\noindent
\textbf{LLM-related limitations.} 
Our approach inherits well-known limitations of LLMs, including hallucinations and non-deterministic behavior. These aspects may lead to spurious taint flows or inconsistent results across runs, making the analysis harder to reproduce and less reliable in security-critical settings. To partially mitigate this issue, we execute each prompt multiple times and retain only the flows that appear in the majority of the runs (see Section~\ref{sec:experimentalsetup}).

\noindent
\textbf{Baseline comparison.}
In this preliminary study, we intentionally compare against the default configuration of FlowDroid to reflect a common usage scenario in which the tool is employed as a black-box taint analyzer. While the literature proposes several extensions targeting specific challenges (e.g., IccTA~\cite{li2015iccta} for ICC analysis), such approaches typically require additional configuration or integration effort beyond the default FlowDroid setup. Investigating whether LLM-enhanced taint analysis approaches can address similar challenges, while maintaining scalability and usability, represents an interesting direction for future work.

%% file: 07_RelatedWork.tex
\section{Related Work}
\label{sec:relatedwork}

Recent work has explored the LLMs adoption for taint analysis. LATTE~\cite{liu2025llm} employs LLMs for static binary taint analysis and vulnerability detection in compiled code. Other works~\cite{li2024iris, ghebremichael2026multi} use LLMs to enhance traditional pipelines by improving taint specifications or assisting vulnerability reasoning. Similarly, J. Ye tal.~\cite{ye2024detecting} applies LLM-based taint reasoning to embedded firmware analysis. However, these approaches do not target Android apps and/or mainly use LLMs as auxiliary steps within existing pipelines, rather than investigating LLM-based reasoning. 
Beyond taint analysis, several recent works have explored leveraging LLMs for Android security tasks, including malware and vulnerability detection~\cite{zhao2025apppoet,qian2025lamd,mathews2024llbezpeky}. 
In contrast, our work explores the use of LLMs to directly reason about end-to-end taint flows in Android apps.

%% file: 08_Conclusion.tex
\section{Conclusion}
\label{sec:conclusion}
This paper presents a preliminary empirical study on the use of LLM reasoning for Android taint analysis. Compared to FlowDroid on \textsc{DroidBench}, our approach achieves promising results, with \BestModel{} reaching an F1-score of \BestFone\ (vs. \FDFone{} for FlowDroid). In particular, the results suggest that LLM-based reasoning can help identify challenging taint flows involving ICC, reflection, and implicit flows.
Rather than replacing traditional static analyzers, LLMs appear promising as a complementary technique for difficult analysis scenarios. Overall, this work provides preliminary evidence to further support research on hybrid taint analysis approaches that combine static analysis with agentic LLM reasoning.

\section*{Data Availability.}
We publicly release all associated resources: 
\begin{center}
   \url{https://github.com/nmiazzomath/Towards-LLM-Enhanced-Android-Taint-Analysis}
\end{center}

\section*{Acknowledgements}
\textbf{LLM usage considerations}: ChatGPT~\footnote{https://chatgpt.com/}, Claude~\footnote{https://claude.ai/} and GitHub Copilot~\footnote{https://github.com/features/copilot} were used for editorial purposes and for generating source code for artifact creation in this work, and all outputs were inspected by the authors to ensure accuracy and originality.
LLMs are also an integral component of our experimental methodology. Because the primary model employed (\texttt{Gemini 3 Flash}) is closed-source, exact reproduction of our results may vary over time. To mitigate this concern, we additionally evaluated an open-source LLM and report their comparative performance.